\documentclass[prd, preprint, nofootinbib]{revtex4}
\usepackage[dvipdfmx]{graphicx}
\usepackage{amsfonts}
\usepackage{latexsym}
\usepackage{amssymb}
\usepackage{epsfig}
\usepackage{comment}
\usepackage{here}	
\usepackage{amsmath}

\def\slashb#1{\not\!\!#1}
\newcommand{\im}[1]{\text{Im}\,#1}

\begin{document}

\title{Quark mass matrices in magnetized orbifold models \\ 
with 
localized Fayet-Iliopoulos terms}

\author{Hiroyuki Abe$^{1}$, Tatsuo Kobayashi$^{2}$, Shintaro~Takada$^{2}$, Shio Tamba$^{2}$, and Takuya H. Tatsuishi$^{2}$}
\affiliation{${}^{1}$Department of Physics, Waseda University, 
Tokyo 169-8555, Japan}
 \affiliation{
${}^{2}$Department of Physics, Hokkaido University, Sapporo 060-0810, Japan}



\begin{abstract}
We study magnetized orbifold models.
We assume the localized Fayet-Iliopoulos terms and the corresponding 
gauge background.
Such terms lead to strong localization of zero-mode wavefunctions.
In this setup, we compute quark mass matrices.
\end{abstract}

\pacs{}
\preprint{EPHOU-18-008}
\preprint{WU-HEP-18-07}

\vspace*{3cm}
\maketitle



\section{Introduction}

Superstring theory is a promising candidate for unified theory of all the interactions including gravity, and  
quarks and leptons as well as Higgs fields.
In addition to our four-dimensional (4D) spacetime, superstring theory predicts extra six dimensional space, 
which should be compact.
Thus, extra dimensional models are well-motivated.

The standard model is the chiral theory.
The torus compactification is one of the 
simplest compactification, but the torus compactification 
leads to a 4D non-chiral low-energy effective field theory.
Thus, the simple torus compactification is not realistic.
How to derive a chiral theory is a key point when we start from extra dimensional field theory and 
superstring theory.

The torus compactification with magnetic flux is a simple way to derive a 4D chiral theory from 
extra dimensional field theory and superstring theory \cite{Bachas:1995ik,Berkooz:1996km,Blumenhagen:2000wh,Angelantonj:2000hi}.
The number of zero-modes, i.e., the generation number is determined by 
the magnitude of magnetic flux.
Their zero-mode wavefunctions are quasi-localized around points different from each other in the compact space.
Such behavior can lead to suppressed couplings, which would be useful to explain 
quark and lepton masses and mixing angles.
Indeed, in Ref.~\cite{Cremades:2004wa}, Yukawa couplings were computed.
In addition, higher order couplings were computed in Ref.~\cite{Abe:2009dr}.
Quark and lepton mass matrices were also discussed.
(See e.g. Refs.~\cite{Abe:2012fj,Abe:2016eyh}.)

The orbifold compactification with magnetic flux is also interesting \cite{Abe:2008fi}.
Orbifolding can project out the adjoint matter fields, i.e., open string moduli.\footnote{
This aspect corresponds to  the T-dual picture of intersecting D-brane models on orbifolds \cite{Blumenhagen:2005tn}. }
Also, the number of zero-modes and their wavefunctions on orbifolds with magnetic fluxes are 
different from those on torus compactification with magnetic flux \cite{Abe:2008fi,Abe:2013bca,Kobayashi:2017dyu}.
Thus, the orbifold compactification makes model building rich.
For example, realization of quark and lepton masses and mixing angles as well as CP phases was studied 
\cite{Abe:2008sx,Abe:2014vza,Kobayashi:2015siy,Matsumoto:2016okl,Fujimoto:2016zjs,Kobayashi:2016qag}.
However, such realization is still a challenging issue.

In addition, we can assume localized operators on orbifold fixed points 
\cite{Buchmuller:2015eya,Buchmuller:2015jna,Buchmuller:2017vho,Ishida:2017avx,Abe:2018qbp}.
That also makes model building more rich.
In Ref.~\cite{Lee:2003mc},  Fayet-Iliopoulos (FI) terms \cite{Fayet:1974jb}, which are localized on the fixed points were studied on the $T^2/Z_2$ orbifold 
compactification (without magnetic flux).
Although the zero-mode profile is flat without localized FI terms, zero-modes are localized around orbifold fixed points because of 
FI terms.
This behavior would be interesting from the phenomenological viewpoint.

In this paper, we study the $Z_2$ orbifold compactification with magnetic flux and localized FI terms.
Localized FI terms drastically change the zero-mode profiles from those without FI terms.
That would also change the pattern of fermion mass matrices.
In this setup, we investigate the realization of quark masses and mixing angles.

This paper is organized as follows.
In Sec. \ref{sec:setup}, we explain our setup.
In Sec. \ref{sec:quark-mass}, we examine quark masses and mixing angles 
in our model.
Sec. \ref{sec:conclusion} is conclusion and discussion.
In Appendix \ref{sec:append}, we show wavefunctions explicitly, which are used in our analysis.

\section{Magnetized orbifold models with localized FI terms}
\label{sec:setup}

In this section, we explain our setup and zero-mode profiles.
We implicitly assume that our models are supersymmetric.
For example, when the compact dimensions is six (four), 
the compact space is the tensor product of three (two) two-dimensional (2D) spaces.
We assume that magnetic fluxes are set such that 4D N=1 supersymmetry is unbroken.
However, the flavor structure, i.e. the difference between fermion flavors 
originates from one of 2D compact spaces.
Hence, we concentrate on the 2D compact space such as 
$T^2$ and $T^2/Z_2$, because this part is important to realize 
Yukawa matrices.
Then, we study the zero-mode wavefunctions and Yukawa couplings 
on the 2D space.

\subsection{Torus compactification with magnetic flux}

Here, we briefly review on the torus model with magnetic flux \cite{Cremades:2004wa}. 
We use the complex coordinate $z= x +\tau y$ instead of the real coordinates $(x,y)$, 
where $\tau$ is a complex structure modulus.
By use of the complex coordinate, the metric is written by 
$ds^2 = g_{\alpha \beta}dz^{\alpha} d\bar{z}^{\beta} $,
\begin{equation}
g_{\alpha \beta} = \left(
\begin{array}{cc}
g_{zz} & g_{z \bar{z}} \\
g_{\bar{z} z} & g_{\bar{z} \bar{z}}
\end{array}
\right) = (2\pi R)^2 \left(
\begin{array}{cc}
0 & \frac{1}{2} \\
\frac{1}{2} & 0
\end{array}
\right) .
\end{equation}
To realize the $T^2$, we identify $z \sim z +1$ and $z \sim z + \tau$.

We consider the U(1) theory with the following magnetic flux, 
\begin{equation}
F = i\frac{\pi M}{\im{\tau}}  (dz \wedge d\bar{z}),
\end{equation}
where $M$ must be quantized to be integer.
This flux can be obtained by the following vector potential,
\begin{equation}
A(z) = \frac{\pi M}{\im{\tau}} \im{(\bar{z}dz)},
\end{equation}
in a certain gauge.

Here, we study the spinor field $\psi(z,\bar z)$ with U(1) charge $q$ on $T^2$, 
which have two components,
\begin{equation}
\psi(z,\bar{z}) = \left(
\begin{array}{c}
\psi_+ \\
\psi_- 
\end{array}
\right).
\end{equation}
Then, we examine the zero-mode equation,
\begin{equation}
i \slashb{D} \psi = 0,
\end{equation} 
which can be written in components,
\begin{equation}
D \psi_+ = 0, \ \ \ D^{\dagger} \psi_- =0,
\end{equation}
where 
\begin{equation}
D^{\dagger} \equiv \partial - q \frac{\pi M}{2 \im{\tau}} \bar{z}, \qquad
D \equiv \bar{\partial} + q \frac{\pi M}{2 \im{\tau}} z .
\end{equation}
Also, they must satisfy the following boundary condition, 
\begin{eqnarray}
\label{eq:z+1}
\psi_\pm(z+1) &=& e^{iq\phi_1(z)}\psi_\pm(z) = \exp \left\{ i \frac{\pi qM}{\im{\tau}} \im{z} \right\} \psi_\pm(z), \\
\label{eq:z+tau}
\psi_\pm(z + \tau) &=& e^{iq\phi_2(z)}\psi_\pm(z) = \exp \left\{ i \frac{\pi qM}{\im{\tau}} \im{\bar{\tau}z} \right\} \psi_\pm(z) .
\end{eqnarray}

Either $\psi_+$ or $\psi_-$ has zero-mode solutions exclusively when $qM \neq 0$.
That is, for $qM >0$ ($qM<0$), $\psi_+$ ($\psi_-$) has $|qM|$ solutions, 
while  $\psi_-$ ($\psi_+$) has no zero-modes.
The number $|qM|$ would correspond to the generation number.
Their zero-mode profiles for $qM > 0$ are given by 
\begin{equation}
\psi^{j,qM}(z) = \mathcal{N} e^{i\pi  qM z \frac{\im{z}}{\im{\tau}}} \cdot \vartheta \left[
\begin{array}{c}
\frac{j}{qM} \\
0
\end{array}
\right] \left( qM z, qM\tau \right),
\end{equation}
with $j=0,1,\cdots, (M-1)$, 
where $\vartheta$ denotes the Jacobi theta function, 
\begin{equation}
\vartheta \left[
\begin{array}{c}
a \\
b
\end{array}
\right] (\nu, \tau) = \sum_{l \in {\bf Z}} e^{\pi i (a+l)^2 \tau} e^{2 \pi i (a+l)(\nu+b)} .
\end{equation}
Here, $\mathcal{N}$ denotes  the normalization factor given by 
\begin{equation}
\label{eq:normalization}
\mathcal{N} = \left( \frac{2\im{\tau} qM}{\mathcal{A}^2} \right)^{1/4}, 
\end{equation}
with $\mathcal{A}= 4 \pi^2 R^2 \im{\tau}$.
The scalar fields have the same wavefunctions as the spinor fields.

\subsection{$T^2/Z_2$ orbifold}

The $T^2/Z_2$ orbifold can be constructed from $T^2$ by  
identifying $z \sim -z$ through the $Z_2$ twist.
For simplicity, we set $\tau = i{\rm Im}\tau$.
 There are four fixed points on $T^2/Z_2$, i.e.,
\begin{equation}
z_I = 0, \qquad \frac12, \qquad \frac{i}{2}{\rm Im}\tau, \qquad \frac12 +  \frac{i}{2}{\rm Im}\tau.
\end{equation}
 
By the $Z_2$ twist $z\rightarrow -z$, the zero-modes can be classified into 
the $Z_2$ even and odd modes.
The zero-modes $\psi^{j,qM}$ satisfy the following relation, 
\begin{equation}
\label{eq:zto-z}
\psi^{j,qM}(-z) = \psi^{qM-j,qM}(z). 
\end{equation}
Note that $\psi^{0,qM}(z)$ is invariant under the $Z_2$ twist, 
and $\psi^{qM/2,qM}(z)$ is also invariant under the $Z_2$ twist 
when $qM$ is even.
Thus, the $Z_2$ even modes are 
\begin{eqnarray}
\label{eq:Z2-wf+}
\Theta^{j,qM}_{+}(z) &=& \frac{1}{\sqrt{2}} \left( \psi^{j,qM}(z) + \psi^{qM-j,qM}(z) \right),
\end{eqnarray}
for $j \neq 0, qM/2$, in addition to 
\begin{equation}
\Theta^{0,qM}_{+}(z) = \psi^{0,qM}(z),
\end{equation}
\begin{equation}
\Theta^{qM/2,qM}_{+}(z) = \psi^{qM/2,qM}(z),
\end{equation}
only if $qM$ is even.
Similarly, the $Z_2$ odd modes are obtained by 
\begin{eqnarray}
\label{eq:Z2-wf-}
\Theta^{j,qM}_{-}(z) &=& \frac{1}{\sqrt{2}} \left( \psi^{j,qM}(z) - \psi^{qM-j,qM}(z) \right).
\end{eqnarray}
The numbers of zero-modes are shown in Table  \ref{tab:Z2}.
The three generations can be obtained as $Z_2$ even (odd) modes for $qM=4,5$  ($qM=7,8$) \cite{Abe:2008sx}.
When we include non-vanishing Wilson lines, the numbers of $Z_2$ even and odd zero-modes change \cite{Abe:2013bca}.
Here, we restrict ourselves to vanishing Wilson lines.

\begin{table}[h]
\begin{tabular}{|c|c|c|} \hline
$qM$ & $2n$ & $2n+1$ \\ \hline \hline
$Z_2$ even & $n+1$ & $n+1$ \\  \hline 
$Z_2$ odd  & $n-1$ &  $n$   \\  \hline
\end{tabular}
\caption{The numbers of $Z_2$ even and odd zero-modes.}
\label{tab:Z2}
\end{table}

\subsection{Localized FI terms without bulk magnetic flux}

Here, we review briefly on the FI terms localized on the orbifold fixed points 
with no bulk magnetic flux \cite{Lee:2003mc}.
The localized FI terms such as
\begin{equation}
\xi = \sum_{I = {\rm f.p.}} (\xi_I + \xi''(\partial \bar \partial))\delta^2(z-z_I) ,
\end{equation}
were studied in Ref.~\cite{Lee:2003mc}, 
where these FI terms were induced by radiative corrections due to 
bulk and brane modes.\footnote{
Similarly, the FI terms are generated by radiative corrections on the $S^1/Z_2$ orbifold, 
and wavefunction profiles are strongly affected by FI terms to be (quasi-)localized.
See, e.g. Refs.~\cite{Ghilencea:2001bw,GrootNibbelink:2002wv,Abe:2002ps}.}
Here, f.p. means that the summation is taken over the fixed points.
Also, it was shown that under the condition of unbroken supersymmetry 
the localized FI terms correspond to the flux,
$F^{(\xi)} = \xi$, which is obtained by the vector potential $A^{(\xi)}$.
Then, the solution of the zero-mode equation, 
\begin{equation}
(\bar{\partial} + q A^{(\xi)} )\psi_+ = 0,
\end{equation}
was studied.
Without $A^{(\xi)}$, the zero-mode profile is constant.
However, with non-vanishing $A^{(\xi)}$, zero-mode profiles are localized around fixed points, 
depending on $q\xi^I$.
Thus, the localized FI terms lead to a strong localization of zero-mode wavefunctions.
Explicit solutions for the above zero-mode equations were shown in Ref.~\cite{Lee:2003mc}, 
\begin{equation}
\phi^{(\xi)} = \prod_{I = {\rm f.p.}}|\vartheta_1(z-z_I|\tau)|^{q\xi_I/(2\pi)} 
\times 
{\rm exp}\left[ q\xi^{''}/R^2\delta^2(z-z_I) +\cdots \right].
\end{equation}
Their wavefunctions can be singular at the fixed points.
A certain regularization was also shown  in Ref.~\cite{Lee:2003mc}.
In this paper, we will assume the same form of the gauge background $A^{(\xi)}$.

\subsection{Magnetized orbifold with localized FI terms}

Here, we explain our setup.
We consider the $T^2/Z_2$ orbifold compactifictaion with bulk magnetic flux and 
FI terms localized on the fixed points.
In other words, we assume the vector potential $A^{(\xi)}$, which  corresponds effectively to the localized FI terms 
in the previous section as Ref.~\cite{Lee:2003mc}.
Then, the zero-mode equation, e.g. for $\psi_+$ is written by 
\begin{equation}
( \bar{\partial} + q \frac{\pi M}{2 \im{\tau}} z + q A^{(\xi)}) \psi_+ =0.
\end{equation}
When $A^{(\xi)} =0$, the solutions on the orbifold are obtained as $\Theta_{\pm}^{j,qM}$.
On the other hand, when $qM=0$, the solution is $\phi^{(\xi)}$.
Then, the solutions of the above zero-mode equation can be written by 
$\Theta_{\pm}^{j,qM} \phi^{(\xi)}$.
However, $\phi^{(\xi)}$ has a singular behavior on the fixed points.
We need some regularization to obtain finite results.

For example, the Yukawa couplings can be computed by overlap integral of the 
wavefunctions such as 
\begin{equation}
Y_{jk\ell} = g \int d^2z ~(\Theta_{\pm}^{j,qM} \phi) (\Theta_{\pm}^{k,q'M'} \phi) (\Theta_{\pm}^{\ell,q''M''} \phi),
\end{equation} 
where $g$ is a coupling in higher dimensional field theory.
In order to derive a finite result on $Y_{jk\ell}$, we need some regularization of $\phi^{(\xi)}$.\footnote{
See for a explicit form of regularization Ref.~\cite{Lee:2003mc}.}
In any regularization, $\phi^{(\xi)}$ would have a huge value around the fixed points, while 
 its value in the bulk except fixed points is suppressed very much compared with a huge value at fixed points.
 Here, instead of using an explicit regularization, we make Ansatz such that 
 the Yukawa couplings can be approximately computed by wavefunctions on fixed points,
 \begin{equation}
Y_{jk\ell} =  \sum_{I={\rm f.p.}}c_I\Theta_{\pm}^{j,qM}(z_I) \Theta_{\pm}^{k,q'M'} (z_I) \Theta_{\pm}^{\ell,q''M''}(z_I ).
\end{equation} 
Here, $c_I$ would depend on our regularization scheme and  the parameters, $qM,q'M',q''M''$ and $\xi_I$.
We use $c_I$ in order to parametrize our ignorance.

Note here that $\Theta_-^{j,qM}(z=0) =0$, because of $\Theta_-^{j,qM}(z) =- \Theta_-^{j,qM}(z)$.
Similarly, we can show that wavefunctions of $Z_2$ odd modes vanish at all of the fixed points 
except the fixed point $z_I=1/2 + i {\rm Im}\tau/2$ 
by use of the boundary conditions (\ref{eq:z+1}) and (\ref{eq:z+tau}), i.e.,
\begin{equation}
\Theta_-^{j,qM}(z=z_I) =0.
\end{equation}
Also, at the fixed point $z_I=1/2 + i {\rm Im}\tau/2$ the wavefunction vanishes 
when $\tau = i {\rm Im} \tau$ and $M=$ even.
Thus, we obtain trivial results for the $Z_2$ odd modes in most of models.
We concentrate on the $Z_2$ even modes.

When one of three fields, say $ \Theta_{+}^{\ell,q''M''}(z_I )$ corresponds 
to the Higgs field in the above Yukawa coupling, and it develops its vacuum expectation value, $v$, 
fermion masses are obtained by 
 \begin{equation}
m_{jk} =  v\sum_{I={\rm f.p.}}c_I\Theta_{+}^{j,qM}(z_I) \Theta_{+}^{k,q'M'} (z_I) \Theta_{+}^{\ell,q''M''}(z_I ).
\end{equation} 
Many models lead to multi Higgs fields.(See, e.g. \cite{Abe:2008sx,Abe:2015yva}.)
The standard-model Higgs field would correspond to their linear combination, 
and the other would gain mass terms at some stage.
In such multi-Higgs models, fermion masses could be written by 
 \begin{equation}
m_{jk} =  \sum_{I={\rm f.p.}}c_I\Theta_{+}^{j,qM}(z_I) \Theta_{+}^{k,q'M'} (z_I)  \left(\sum_\ell v^\ell \Theta_{+}^{\ell,q''M''}(z_I ) \right).
\end{equation} 
However, these masses can be parametrized by 
 \begin{equation}
 \label{eq:mass-Ansatz}
m_{jk} =  \sum_{I={\rm f.p.}}m_I~\Theta_{+}^{j,qM}(z_I) \Theta_{+}^{k,q'M'} (z_I) .
\end{equation} 
Here, four parameters, $m_I$ include our ignorance in $c_I$ and details in the Higgs sector.\footnote{A similar mass matrix is obtained for multiple Higgs generations from localized $\mu$-terms at the fixed points on magnetized orbifolds~\cite{Abe:2018qbp}.}
In the next section, we use this Ansatz to study quark masses and mixing angles.
Note that if only one of $m_I$ is non-vanishing and the other vanishes, 
the mass matrix $m_{ij}$ has rank 1.
Thus, two or more $m_I$'s must be non-vanishing.

\section{Quark mass matrices}
\label{sec:quark-mass}

Here we study quark masses and mixing angles by using Ansatz (\ref{eq:mass-Ansatz}) in the previous section.
Recall that for simplicity, we set $\tau$ to be pure imaginary,  i.e. $\tau =i {\rm Im} \tau $.
The number of $Z_2$ even zero-modes is equal to three only if $qM=4$ and 5.
Their wavefunctions are explicitly shown in Appendix \ref{sec:append}.

We do not construct a model explicitly.
However, we assume that the quark doublets and the up-sector of right-handed quarks 
correspond to $qM=4$, and the down-sector of right-handed quarks correspond to 
$qM=5$.
We assign the first, second and third families to 
$\Theta^{2,qM}_+$, $\Theta^{1,qM}_+$, and $\Theta^{0,qM}_+$,  respectively.
We also introduce the up-sector and down-sector of Higgs fields, which are independent of 
each other.
Then, following the Ansatz (\ref{eq:mass-Ansatz}) and wavefunctions shown in Appendix \ref{sec:append}, 
the up-sector quark mass matrix can be written 
\begin{equation}
M^{(u)}=M^{(u)}_{0,0}+M^{(u)}_{1/2,0}+M^{(u)}_{0,1/2}+M^{(u)}_{1/2,1/2},
\end{equation}
where 
\begin{equation}
M^{(u)}_{0,0} = m_{0,0}^{(u)}\left(
\begin{array}{ccc}
4 e^{-2\tau'}  &2\sqrt{2} e^{-5\tau'/4} & 2 e^{-\tau'}   \\
2\sqrt{2} e^{-5\tau'/4} &  2 e^{-\tau'/2} & \sqrt{2} e^{-\tau'/4} \\
 2 e^{-\tau'}  & \sqrt{2} e^{-\tau'/4}  & 1
\end{array}\right),
\end{equation}
\begin{equation}
M^{(u)}_{1/2,0} = m_{0,0}^{(u)}\left(
\begin{array}{ccc}
4 e^{-2\tau'}  &-2\sqrt{2} e^{-5\tau'/4} & 2 e^{-\tau'}   \\
-2\sqrt{2} e^{-5\tau'/4} &  2 e^{-\tau'/2} & -\sqrt{2} e^{-\tau'/4} \\
 2 e^{-\tau'}  & -\sqrt{2} e^{-\tau'/4}  & 1
\end{array}\right),
\end{equation}
\begin{equation}
M^{(u)}_{0.1/2} = m_{0,1/2}^{(u)}\left(
\begin{array}{ccc}
1  &\sqrt{2} e^{-\tau'/4} & 2 e^{-\tau'}   \\
\sqrt{2} e^{-\tau'/4} &  2 e^{-\tau'/2} & 2\sqrt{2} e^{-5\tau'/4} \\
 2 e^{-\tau'}  & 2\sqrt{2} e^{-5\tau'/4}  & 4 e^{-2\tau'}
\end{array}\right),
\end{equation}
\begin{equation}
M^{(u)}_{1/2,1/2} = m_{1/2,1/2}^{(u)}\left(
\begin{array}{ccc}
1  &-\sqrt{2} e^{-\tau'/4} & 2 e^{-\tau'}   \\
-\sqrt{2} e^{-\tau'/4} &  2 e^{-\tau'/2} & -2\sqrt{2} e^{-5\tau'/4} \\
 2 e^{-\tau'}  & -2\sqrt{2} e^{-5\tau'/4}  & 4 e^{-2\tau'}
\end{array}\right),
\end{equation}
up to the normalization factor $\mathcal{N}$.
Here we define $\tau' =\pi {\rm Im}\tau$.
Similarly, the down-sector quark mass matrix is written by 
\begin{equation}
M^{(d)}=M^{(d)}_{0,0}+M^{(d)}_{1/2,0}+M^{(d)}_{0,1/2}+M^{(d)}_{1/2,1/2},
\end{equation}
where 
\begin{equation}
M^{(d)}_{0,0} = m_{0,0}^{(d)}\left(
\begin{array}{ccc}
2\sqrt{2} e^{-9\tau'/5}  &2\sqrt{2} e^{-6\tau'/5} & 2 e^{-\tau'}   \\
2 e^{-21\tau'/20} &  2 e^{-9\tau'/20} & \sqrt{2} e^{-\tau'/4} \\
 \sqrt{2} e^{-4\tau'/5}  & \sqrt{2} e^{-\tau'/5}  & 1
\end{array}\right),
\end{equation}
\begin{equation}
M^{(d)}_{1/2,0} = m_{1/2,0}^{(d)}\left(
\begin{array}{ccc}
2\sqrt{2} e^{-9\tau'/5}  &-2\sqrt{2} e^{-6\tau'/5} & 2 e^{-\tau'}   \\
-2 e^{-21\tau'/20} &  2 e^{-9\tau'/20} & -\sqrt{2} e^{-\tau'/4} \\
 \sqrt{2} e^{-4\tau'/5}  & -\sqrt{2} e^{-\tau'/5}  & 1
\end{array}\right),
\end{equation}
\begin{equation}
M^{(d)}_{0,1/2} = m_{0,1/2}^{(d)}\left(
\begin{array}{ccc}
\sqrt{2} e^{-\tau'/20}  &\sqrt{2} e^{-9\tau'/20} & 2 e^{-5\tau'/4}   \\
2 e^{-6\tau'/20} &  2 e^{-14\tau'/20} & 2\sqrt{2} e^{-3\tau'/2} \\
 2\sqrt{2} e^{-21\tau'/20}  & 2\sqrt{2} e^{-29\tau'/20}  & 4 e^{-9\tau'/4} 
\end{array}\right),
\end{equation}
up to the normalization factor $\mathcal{N}$.
In addition, we have $M^{(d)}_{1/2,1/2}=0\times m^{(d)}_{1/2,1/2}$ for ${\rm Re}\tau =0$.
For ${\rm Re}\tau \neq 0$, all the entries of the matrix $M^{(d)}_{1/2,1/2}$ are non-vanishing, 
and their absolute values are similar to $M^{(d)}_{0,1/2}$.

We have nine parameters, $m^{(u,d)}_{0,0}, m^{(u,d)}_{1/2,0},m^{(u,d)}_{0,1/2}, m^{(u,d)}_{1/2,1/2}$, and 
${\rm Im}\tau$, and this number of free parameters is 
enough to fit them to experimental data of six quarks masses and three mixing angles.
When we include non-vanishing ${\rm Re} \tau$, we can also fit the CP phase.
Thus, we do not examine detailed fitting, but we study order estimation.
Note that $M^{(u)}_{0,0}$ ($M^{(u)}_{0,1/2}$) is very similar to $M^{(u)}_{1/2,0}$ ($M^{(u)}_{1/2,1/2}$) , 
and $M^{(d)}_{0,0}$ ($M^{(d)}_{0,1/2}$) is very similar to $M^{(d)}_{1/2,0}$ ($M^{(d)}_{1/2,1/2}$).\footnote{
When ${\rm Re}\tau \neq 0$, $M^{(d)}_{0,1/2}$ is similar to $M^{(d)}_{1/2,1/2}$.}
For simple estimation, we consider the parameter region,
\begin{eqnarray} 
\label{eq:parameter}
m^{(u)} &\sim& m^{(u)}_{0,0} + m^{(u)}_{1/2,0} \sim  m^{(u)}_{0,0} - m^{(u)}_{1/2,0},  \nonumber \\
\rho^{(u)}m^{(u)} &\sim& m^{(u)}_{0,1/2} + m^{(u)}_{1/2,1/2} \sim  m^{(u)}_{0,1/2} - m^{(u)}_{1/2,1/2},  \nonumber \\
m^{(d)} &\sim& m^{(d)}_{0,0} + m^{(d)}_{1/2,0} \sim  m^{(d)}_{0,0} - m^{(d)}_{1/2,0},   \\
\rho^{(d)}m^{(d)} &\sim& m^{(d)}_{0,1/2}. \nonumber
\end{eqnarray}
Then, the quark mass matrices can be written by 
\begin{equation}
\frac{M^{(u)}}{m^{(u)}} \sim \left(
\begin{array}{ccc}
4 e^{-2\tau'} + \rho^{(u)}
&\sqrt{2} (2e^{-5\tau'/4}+ \rho^{(u)}  e^{-\tau'/4}  )
& 2 e^{-\tau'} (1 + \rho^{(u)})  \\
\sqrt{2} (2e^{-5\tau'/4}+ \rho^{(u)}  e^{-\tau'/4} ) 
&  2 e^{-\tau'/2}(1+ \rho^{(u)}  )
& \sqrt{2}( e^{-\tau'/4}+2 \rho^{(u)} e^{-5\tau'/4} )  \\
 2 e^{-\tau'}(1 + \rho^{(u)})  
& \sqrt{2} (e^{-\tau'/4} +2 \rho^{(u)}  e^{-5\tau'/4} ) 
 & 1+ 4\rho^{(u)}  e^{-2\tau'}
\end{array}\right),
\end{equation}
\begin{equation}
\frac{M^{(d)}}{m^{(d)}} \sim \hskip -.2cm \left(
\begin{array}{ccc}
\sqrt{2} (2e^{-9\tau'/5}  +\rho^{(d)}e^{-\tau'/20})
&\sqrt{2} (2e^{-6\tau'/5} +\rho^{(d)}e^{-9\tau'/20})
& 2 (e^{-\tau'}  +\rho^{(d)}e^{-5\tau'/4} ) \\
2 (e^{-21\tau'/20} +\rho^{(d)}e^{-6\tau'/20})
&  2 (e^{-9\tau'/20} +\rho^{(d)}e^{-14\tau'/20})
& \sqrt{2}( e^{-\tau'/4} +2\rho^{(d)} e^{-3\tau'/2})  \\
 \sqrt{2} (e^{-4\tau'/5}  +2\rho^{(d)} e^{-21\tau'/20})
& \sqrt{2} (e^{-\tau'/5}  +2\rho^{(d)}e^{-29\tau'/20})
& 1+4\rho^{(d)}e^{-9\tau'/4} 
\end{array}\right).
\end{equation}

It is very straightforward to realize the top and bottom quark masses.
Hence, we try to fit our three parameters, $\rho^{(u),(d)}$ and $\tau'$ to 
mass ratios, $m_c/m_t$, $m_u/m_t$, $m_s/m_b$, $m_d/m_b$ and mixing angles, 
seven observables.
The experimental values of the Cabibbo-Kobayashi-Maskawa matrix are 
\begin{equation}
|V_{\rm CKM}| = \left(
\begin{array}{ccc}
0.97 & 0.23 & 0.0035 \\
0.23 & 0.97 & 0.041 \\
0.0087 & 0.040 & 1.0
\end{array}\right).
\end{equation}
For example, at 1 TeV , rations of running masses are obtained as 
\cite{Xing:2007fb,Antusch:2013jca},
\begin{eqnarray}
& & \frac{m_c}{m_t} = 3.5 \times 10^{-3}, \qquad \frac{m_u}{m_t} = 7.3 \times 10^{-6}, \nonumber \\
& & \frac{m_s}{m_b} = 1.9 \times 10^{-2}, \qquad \frac{m_d}{m_b} = 1.0 \times 10^{-3}.
\end{eqnarray}

We concentrate on the parameter region, $|\rho^{(u,d)}|  \ll 1$.
First, we study the mass matrices of the second and third generations, 
which are written for $|\rho^{(u,d)}|  \ll 1$ as 
\begin{equation}
\frac{M^{(u)}}{m^{(u)}} \sim \left(
\begin{array}{cc}
 2 e^{-\tau'/2}
& \sqrt{2} e^{-\tau'/4}  \\
\sqrt{2} e^{-\tau'/4} 
 & 1
\end{array}\right),
\qquad 
\frac{M^{(d)}}{m^{(d)}} \sim  \left(
\begin{array}{cc}
 2 e^{-9\tau'/20} 
& \sqrt{2} e^{-\tau'/4}  \\
 \sqrt{2} e^{-\tau'/5}  
& 1
\end{array}\right).
\end{equation}
These mass matrices include only one free parameter $\tau'$, and are 
very predictable.
For example, we take $\sqrt{2}e^{-\tau'/4}=0.08$.
Then, we obtain 
\begin{equation}
 \frac{m_c}{m_t} = 6.4 \times 10^{-3},  \qquad 
 \frac{m_s}{m_b} = 1.1 \times 10^{-2}, \qquad V_{cb} = 0.08.
\end{equation} 
These orders are consistent with experimental values.

Next, we examine the other mixing angles and the mass ratios, 
$m_u/m_t$ and $m_d/m_b$.
For example, when we take $\rho^{(u)} \sim 7.3 \times 10^{-3}$, we can 
realize the experimental values $m_u/m_t$.
Also, when we take $\rho^{(d)} ={\cal O}(0.1)$, 
we can  realize the experimental order of $V_{us}$.
However, for this value of  $\rho^{(d)} $, we have a large ratio, 
$m_d/m_b={\cal O}(0.1)$.
On the other hand, when take  $\rho^{(d)} ={\cal O}(0.001)$, 
we can realize the experimental value $m_d/m_b$, 
but we have a small value $V_{us} = {\cal O}(0.001)$.
Thus, we can realize most of experimental values by 
the simple parameter region (\ref{eq:parameter}), 
although there is a tension between $m_d/m_b$ and  $V_{us}$.
However, by tuning $m^{(d)}_{0,1/2}$ and $m^{(d)}_{1/2,1/2}$ as well as ${\rm Re}\tau$, 
we can realize both  $m_d/m_b$ and  $V_{us}$.


\section{Conclusion}
\label{sec:conclusion}
\label{sec:conclusion}

We have studied magnetized orbifold models.
We have assumed the FI terms localized at fixed points and 
the corresponding gauge background.
Such terms lead to strong localization of zero-mode wavefunctions.
We have computed quark mass matrices by parameterizing 
detail of models and our ignorance.
The forms of quark mass matrices are quite simple, 
but we can fit the experimental data $m_c/m_t$, $m_s/m_b$, $V_{cb}$ roughly by 
just one parameter, $\tau'$.
We can also realize $m_u/m_t$.
However, there is a tension between $m_d/m_b$ and $V_{us}$ in the simple parameter region, 
although we can tune parameters to realize both $m_d/m_b$ and $V_{us}$.

Similarly, we can discuss the lepton sector.
For the mass matrix of charged leptons, the analysis is similar and 
straightforward.
We can realize charged lepton masses.
For the neutrino masses and mixing angles, 
it is an important issue how to derive neutrino masses.
For example, right-handed Majorana neutrino masses can be generated 
on magnetized orbifold models by D-brane instanton effects \cite{Kobayashi:2015siy}.
It would be important to study such D-brane instanton effects 
under the background corresponding the localized FI terms.

In this paper, we have concentrated on the $T^2/Z_2$ orbifold without 
discrete Wilson lines.
It would be interesting to extend our analysis to the 
$T^2/Z_2$ orbifold with discrete Wilson lines and other orbifolds 
with discrete Wilson lines \cite{Abe:2013bca}.
The numbers of fixed points on the other orbifolds are different from 
one of the $T^2/Z_2$ orbifold, and their fixed point structures are 
different.
For example, the $T^2/Z_3$ orbifold has three fixed points, and 
the number of free parameters corresponding to $c_I$ and $m_I$ is three.
Hence, it would be intriguing to study the $T^2/Z_3$ orbifold.
We would study elsewhere.


\section*{Acknowledgments}
Authors would like to thank M.Ishida for useful discussions.
H.~A.  is supported in part by JSPS KAKENHI Grant Number JP16K05330.
T.~K.  is supported in part by MEXT KAKENHI Grant Number JP17H05395.
T.~H.~T is supported in part by Grant-in-Aid for JSPS Research 
Fellow (No. 18J11233).

%


\appendix

\section{Wavefunctions}
\label{sec:append}

The number of $Z_2$ even zero-modes is equal to three only if $qM=4$ and 5.
For simplicity, we set $\tau = i {\rm Im}\tau$.
Their wavefunctions are approximated up to the normalization  $\mathcal{N}$ as 
\begin{eqnarray}
\Theta^{0,4}_+(z) &\sim&  1+ 2e^{-4\pi {\rm Im}\tau} + \cdots \qquad {\rm at~~} z=0, \nonumber \\
                   &\sim&   1+ 2e^{-4\pi {\rm Im}\tau} + \cdots \qquad {\rm at~~} z=\frac12,  \nonumber \\
                    &\sim&   2\left( e^{-\pi {\rm Im}\tau} +  e^{-9\pi {\rm Im}\tau} \cdots \right) \qquad {\rm at~~} z=\frac{i}{2}{\rm Im}\tau,\\
                    &\sim&   -2\left(e^{-\pi {\rm Im}\tau} +  e^{-9\pi {\rm Im}\tau} \cdots  \right) \qquad {\rm at~~} z=\frac12+\frac{i}{2}{\rm Im}\tau,  \nonumber
\end{eqnarray}
 for      $\Theta^{0,4}_+(z)$,              
\begin{eqnarray}
\Theta^{1,4}_+(z) &\sim&  \sqrt{2}\left(e^{-(1/4)\pi {\rm Im}\tau} + e^{-(9/4)\pi {\rm Im}\tau} + \cdots \right) \qquad {\rm at~~} z=0,  \nonumber \\
                   &\sim&  \sqrt{2}\left(-e^{-(1/4)\pi {\rm Im}\tau} - e^{-(9/4)\pi {\rm Im}\tau} + \cdots \right)\qquad {\rm at~~} z=\frac12, \nonumber \\
                    &\sim&   \sqrt{2} \left(e^{-(1/4)\pi {\rm Im}\tau} +  e^{-(9/4)\pi {\rm Im}\tau} \cdots \right) \qquad {\rm at~~} z=\frac{i}{2}{\rm Im}\tau,\\
                    &\sim&  \sqrt{2} \left(e^{-(1/4)\pi {\rm Im}\tau} +  e^{-(9/4)\pi {\rm Im}\tau} \cdots \right) \qquad {\rm at~~} z=\frac12+\frac{i}{2}{\rm Im}\tau, \nonumber
\end{eqnarray}
 for      $\Theta^{1,4}_+(z)$,      
\begin{eqnarray}
\Theta^{2,4}_+(z) &\sim&  2\left(e^{-\pi {\rm Im}\tau} + e^{-9\pi {\rm Im}\tau} + \cdots \right) \qquad {\rm at~~} z=0,  \nonumber \\
                   &\sim&  2\left(e^{-\pi {\rm Im}\tau} + e^{-9\pi {\rm Im}\tau} + \cdots \right)\qquad {\rm at~~} z=\frac12, \nonumber \\
                    &\sim&   1 + 2 e^{-4\pi {\rm Im}\tau} + \cdots \qquad {\rm at~~} z=\frac{i}{2}{\rm Im}\tau,\\
                    &\sim&  -1 -2 e^{-4\pi {\rm Im}\tau} \cdots  \qquad {\rm at~~} z=\frac12+\frac{i}{2}{\rm Im}\tau,  \nonumber
\end{eqnarray}
 for      $\Theta^{2,4}_+(z)$,      

\begin{eqnarray}
\Theta^{0,5}_+(z) &\sim&  1+ 2e^{-5\pi {\rm Im}\tau} + \cdots \qquad {\rm at~~} z=0,  \nonumber \\
                   &\sim&   1-2 e^{-5\pi {\rm Im}\tau} + \cdots \qquad {\rm at~~} z=\frac12, \nonumber  \\
                    &\sim&   2\left( e^{-(5/4)\pi {\rm Im}\tau} +  e^{-(45/4)\pi {\rm Im}\tau} \cdots \right) \qquad {\rm at~~} z=\frac{i}{2}{\rm Im}\tau,\\
                    &=&   0 \qquad {\rm at~~} z=\frac12+\frac{i}{2}{\rm Im}\tau,  \nonumber
\end{eqnarray}
 for      $\Theta^{0,5}_+(z)$, 

\begin{eqnarray}
\Theta^{1,5}_+(z) &\sim&  \sqrt{2}\left(e^{-(1/5)\pi {\rm Im}\tau} + e^{-(16/5)\pi {\rm Im}\tau} + \cdots \right) \qquad {\rm at~~} z=0,  \nonumber \\
                   &\sim&  \sqrt{2}\left(-e^{-(1/5)\pi {\rm Im}\tau} + e^{-(16/5)\pi {\rm Im}\tau} + \cdots \right)\qquad {\rm at~~} z=\frac12, \nonumber \\
                    &\sim&   \sqrt{2} \left(e^{-(9/20)\pi {\rm Im}\tau} +  e^{-(49/20)\pi {\rm Im}\tau} \cdots \right) \qquad {\rm at~~} z=\frac{i}{2}{\rm Im}\tau,\\
                    &=&   0 \qquad {\rm at~~} z=\frac12+\frac{i}{2}{\rm Im}\tau,  \nonumber
\end{eqnarray}
 for      $\Theta^{1,5}_+(z)$,      

\begin{eqnarray}
\Theta^{2,5}_+(z) &\sim&  \sqrt{2}\left(e^{-(4/5)\pi {\rm Im}\tau} + e^{-(9/5)\pi {\rm Im}\tau} + \cdots \right) \qquad {\rm at~~} z=0,  \nonumber \\
                   &\sim&  \sqrt{2}\left(e^{-(4/5)\pi {\rm Im}\tau} - e^{-(9/5)\pi {\rm Im}\tau} + \cdots \right)\qquad {\rm at~~} z=\frac12, \nonumber \\
                    &\sim&   \sqrt{2} \left(e^{-(1/20)\pi {\rm Im}\tau} -  e^{-(81/20)\pi {\rm Im}\tau} \cdots \right) \qquad {\rm at~~} z=\frac{i}{2}{\rm Im}\tau,\\
                    &=&   0 \qquad {\rm at~~} z=\frac12+\frac{i}{2}{\rm Im}\tau,  \nonumber
\end{eqnarray}
 for      $\Theta^{2,5}_+(z)$.  
At $z=\frac12+\frac{i}{2}{\rm Im}\tau$, we have $\Theta^{0,5}_+(z) = \Theta^{1,5}_+(z)= \Theta^{2,5}_+(z) =0$ for ${\rm Re}\tau =0$.
However, when we set ${\rm Re}\tau \neq 0$, we obtain non-vanishing values of $\Theta^{0,5}_+(z), \Theta^{1,5}_+(z),\Theta^{2,5}_+(z) $ 
at the fixed point, $z=\frac12+\frac{\tau}{2}$.



\end{document}